\documentclass[Afour,sagev,times,doublespace]{IEEETran}

\usepackage{moreverb,url}

\usepackage[colorlinks,bookmarksopen,bookmarksnumbered,citecolor=red,urlcolor=red]{hyperref}

\usepackage{graphicx}

\newcommand\BibTeX{{\rmfamily B\kern-.05em \textsc{i\kern-.025em b}\kern-.08em
T\kern-.1667em\lower.7ex\hbox{E}\kern-.125emX}}

\def\volumeyear{2019}

\begin{document}


\title{Visualization in the preprocessing phase: an interview study with enterprise professionals}

\author{\IEEEauthorblockN{Alessandra Milani\textsuperscript{1,3} and}
\and
\IEEEauthorblockN{Fernando Paulovich\textsuperscript{2} and}
\and
\IEEEauthorblockN{Isabel Manssour\textsuperscript{1}}
\\\textsuperscript{1}Pontifical Catholic University of Rio Grande do Sul; \textsuperscript{2}Dalhousie University
\\\textsuperscript{3}alessandra.paz@acad.pucrs.br
}

\maketitle

\begin{abstract}
The current information age has increasingly required organizations to become data-driven. However, analyzing and managing raw data is still a challenging part of the data mining process. Even though we can find interview studies proposing design implications or recommendations for future visualization solutions in the data mining scope, they cover the entire workflow and do not fully focus on the challenges during the preprocessing phase and on how visualization can support it. Moreover, they do not organize a final list of insights consolidating the findings of other related studies. Hence, to better understand the current practice of enterprise professionals in data mining workflows, in particular during the preprocessing phase, and how visualization supports this process, we conducted semi-structured interviews with thirteen data analysts. The discussion about the challenges and opportunities based on the responses of the interviewees resulted in a list of ten insights. This list was compared with the closest related works, improving the reliability of our findings and providing background, as a consolidated set of requirements, for future visualization research papers applied to visual data exploration in data mining. Furthermore, we provide greater details on the profile of the data analysts, the main challenges they face, and the opportunities that arise while they are engaged in data mining projects in diverse organizational areas.
\end{abstract}



\section{Introduction}
\label{intro}

The data-driven society in which we live led us to accumulate massive volumes of data in the most variety of domains. The process of data analysis for knowledge extraction is still a very challenging, laborious activity. During the process of data exploration, data analysts spend most of their time on data preparation activities~\cite{Wiley_Dasu_Johnson_2003}, i.e., the preprocessing phase, when we consider data mining~\cite{Morgan_2011_Han} workflows, such as knowledge discovery in databases (KDD)~\cite{Book_Piateski_1991} or Cross Industry Process for Data Mining (CRISP-DM)~\cite{xxx_shearer_2000}. As examples of the demanding activities that are part of the preprocessing phase, we can list completeness and conformity of data quality, since there is not a single technique or tool to solve all data issues automatically~\cite{ACM_Kandel_2011, XXX_Hellerstein_2008}. Therefore, intense interaction between raw data and data analysts is required to perform the decisions on how to proceed with the data management~\cite{Wiley_Dasu_Johnson_2003, Wiley_Jugulum_2014}.

Consequently, the preprocessing purpose of transforming ``the raw input data into an appropriate format for subsequent analysis''\cite{Book_TAN_2005} may often not be carried out impartially, which means new issues may arise due to the data analysts. For instance, they can update missing values with the mean calculated based on other instances in their dataset instead of the median to avoid outliers or they can even ignore data, e.g., deleting instances due to missing values in a specific attribute, which was supposed to be fixed before proceeding with the data analysis. Thus, no matter how robust the algorithm created for data mining is, if bad data from a source is used or a data manipulation strategy is wrongly selected, it may lead to the identification of wrong patterns and misunderstanding in the final results~\cite{Wiley_Dasu_Johnson_2003}.

Under these circumstances, visualization techniques and visual data exploration could play an important role in data analysis while providing meaningful insights~\cite{IEEE_Oliveira_Levkowitz_2003,Wiley_Jugulum_2014,Book_Ward_2015}. However, most of the visualization studies are concerned with the end of the process when sharing the final results of the analysis. Likewise, we can find interview studies with enterprise professionals proposing design implications~\cite{IEEE_BATCH_2018,IEEE_KANDEL_2012} or recommendations~\cite{IEEE_Alspaugh_2018} for future visualization solutions in the data mining scope, but they cover the entire workflow and do not focus fully in the challenges during the preprocessing phase and on how visualization can support it. Moreover, they do not organize a final list of insights consolidating the findings of other related studies.

In this paper, we aim to gather requirements of how visualization can be used as a powerful tool to be incorporated into the toolkit of the data analysts during the preprocessing phase to foster visual data exploration. We conducted an interview study with thirteen enterprise professionals to investigate their working practices. As a result, we present a consolidated list of ten insights as to how visualization can support the preprocessing activities based on the data analysts’ perspective on data exploration. Furthermore, when analyzing the responses of the interviewees, we provide greater details on the profile of the data analysts, the main challenges they face, and the opportunities that arise while they are engaged in data mining projects in diverse organizational areas.

It is important to highlight that the summarization of practical items, such as ten rules of thumb, provides an overview of the requirements in the preprocessing phase for new visualization efforts, speeding up newcomers’ progress. We also hope it serves as a background for future studies on visualization research applied to data mining, contributing to create awareness of the current gaps and to increase the adoption of visualization techniques as part of the daily practice of data analysts, mainly earlier in their workflow.

The remainder of this paper is structured as follows: Section ``Related Work'' describes the literature review methodology and the interview studies focusing on capturing the experience of data analysts while evaluating design implications in the data mining scope. Subsequently, ``Interview Study'' outlines the procedure developed to perform the interviews, the profile of the participants, and the results and analysis of the interviews. Section ``Insights for New Visualizations'' presents the list of insights resulting from our study. Section ``Discussion and Limitations'' summarizes the discussion of limitations in our study and details the comparative analysis with the related work. Finally, Section ``Conclusion and Future Work'' presents our conclusions and plans for future work.

\section{Related Work}
\label{related_work}

\begin{figure*}[!htb]
   \centering
   \includegraphics[scale=1]{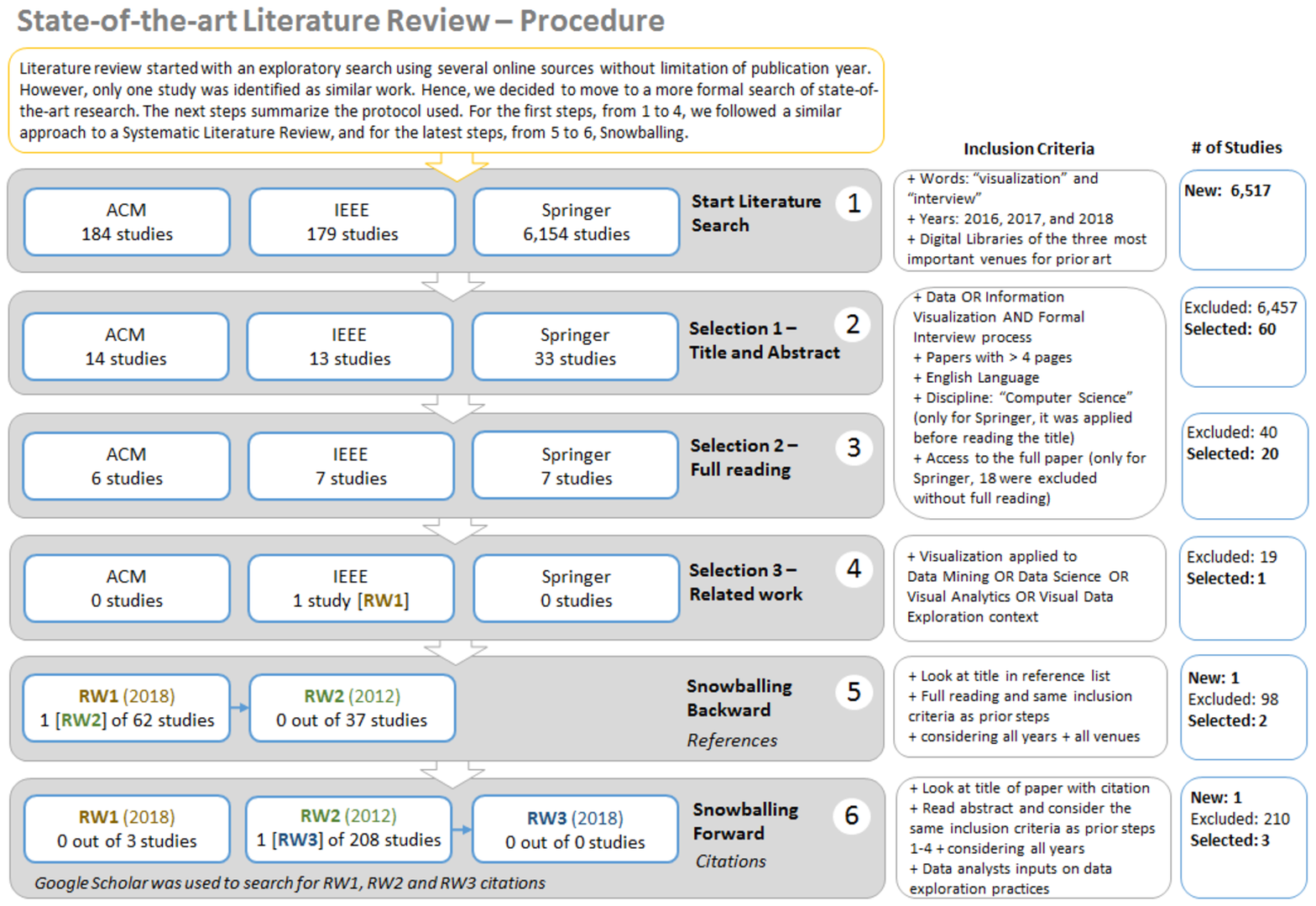}
    \caption{The inclusion criteria for each analyzed study were progressive until Step 4. For Steps 5 and 6, we changed two of the prior inclusion criteria. First, for Step 5, while searching for new references on the list featured in RW1~\cite{IEEE_BATCH_2018}, the year of publication was unlimited, which allowed the selection of RW2~\cite{IEEE_KANDEL_2012}, from 2012. Second, for Step 6, while searching for citations, studies which contributed with the perceptions of professional data analysts on the data exploration process were selected even if the study did not focus on visualization. Hence, a third study was selected, RW3~\cite{IEEE_Alspaugh_2018}. Among the venues for crucial prior studies, three digital libraries, i.e., ACM, IEEE, and Springer, were selected aiming to cover the most relevant journals and conferences in our research scope, in addition to  studies that went through a rigorous review process.}
    \label{fig:LR}
\end{figure*}

We conducted a state-of-the-art literature review to explore interview studies capturing the experience of data analysts while visualizing data during the data mining process. More specifically, we were interested in studies presenting visualization guidelines, challenges, opportunities, or gaps in the preprocessing phase. However, since during the exploratory search for the related work we could not find studies focusing on the preprocessing phase, we then decided to also include studies related to an upper level, e.g., data mining, data analysis, or data science, since their workflows contemplate preprocessing activities.

In brief, Figure~\ref{fig:LR} shows the literature review procedure. Initially, four steps were planned following a systematic literature review process. However, we decided to add two new steps, since up to Step 4 only one study met all the inclusion criteria, presented in Figure~\ref{fig:LR}. Thus, Steps 5 and 6 followed the snowballing search methodology~\cite{ACM_Wohlin_2014}, in an attempt to select additional research, which resulted in a final list of three studies. All these studies presented a discussion on data analysis from the perspective of enterprise professionals and used interviews with semi-structured questionnaires as a data collection instrument. They are referenced in this work as RW1 for Batch and Elmqvist~\cite{IEEE_BATCH_2018}, RW2 for Kandel et al.~\cite{IEEE_KANDEL_2012}, and RW3 for Alspaugh et al.~\cite{IEEE_Alspaugh_2018}.

RW1 developed a variant of contextual inquiry to observe eight data analysts in their work environment. All the participants worked for the U.S. Government in Washington, D.C.. Their experience in data science ranged from four to twenty years. The interview analysis was very detailed, however, the main limitation of the study is the lack of representation of professionals from different sectors. On the contrary, RW2 interviewed 35 enterprise analysts who were working in 25 organizations across a variety of industries. Even though most of the participants were located in Northern California, in the U.S., this scenario brought good coverage of heterogeneous experiences and responses to be analyzed. However, the activities for the preprocessing phase were not fully explored since the study aimed to characterize the space of analytic workflows as a whole.

Even though RW3 did not aim primarily to explore visualization options, its results, based on interviews with thirty data analysts located in the San Francisco Bay Area, in the U.S., were still relevant to us, in particular because they presented an extensive discussion on data exploration practices, which included visualization as a tool.

To summarise, these three studies proposed design implications (RW1 and RW2) or recommendations (RW3) for future tools in data exploration or visual analytics research. Their investigation contributed to identifying challenges, opportunities, and barriers to adopt visualization during exploratory data analyses. Hence, they were used to ratify most of the items included in our final list of insights for new visualizations.

Nevertheless, we can still highlight relevant differences when comparing them with the proposal of our study. First, in our research, we explore aspects to broaden the understanding of how the preprocessing phase is performed in data mining workflows and we instigate the discussion on how visualization could contribute to that process. Furthermore, we go into greater detail concerning the profile of the data analysts, including a description of their work process, details on data type and source, tools and technologies, and strategies for data mining or machine learning in use. Finally, we compiled a more straightforward list of requirements for future visualization solutions in this research area, considering the inputs received by enterprise professionals combined with the review of these three related works.

\section{Interview Study}
\label{interviews}

As a qualitative data collection instrument, we developed a semi-structured questionnaire to guide the interviews with the data analysts. Most of the questions were open-ended in order to capture as much information as possible during the interviews. Some questions covered the participant's profile with a few demographic items. Others were intended to encourage the participants to describe their working practices to provide an overview of their data exploration processes. In addition, some questions were phrased specifically to address the visualization strategies as part of the preprocessing activities. Furthermore, few related works~\cite{IEEE_KANDEL_2012,IEEE_BATCH_2018,IEEE_LAM_2012} were used as reference points during the development of the procedure and the definition of the questions.

\subsection{Participants}

We set as a goal to interview between 10 to 15 data analysts considering the research methods in Human-Computer Interaction~\cite{Book_Lazar_2017}. The participants were recruited based on their engagement with the practice of data mining. We used online platforms, such as Linkedin and Meetup, and our professional network to identify potential participants. We interviewed a total of thirteen professionals, twelve male and one female, with ages ranging from 26 to 42. They were located in three different cities from the same country, but geographically distant. 

Our participants worked in different areas, such as Technology Consulting and Services, Education, Finances, Web Portals, Statistical Consulting, and E-commerce. Twelve of them worked in the private sector, and only one participant had a governmental job. There were three cases where they held positions at the Industry and the Academy at the same time. The range of their company size was significantly wide, from three to close to a hundred thousand collaborators. Their organizational roles varied from Director or Manager (31\%) to Researcher (23\%), but most of them were officially Data Scientists or Data Analysts (46\%). 

The majority of participants (85\%) had received master's degrees in Computer Science, Engineering, Statistics, or Business Administration. One of them completed a Ph.D. program, and three were Ph.D. candidates. Their background during their undergraduate studies included different areas such as Physics, Statistics, Engineering, and Business Administration. However, Computer-Science-related areas were still predominant among this group.

\begin{figure}[!htb]
   \centering
   \includegraphics[scale=0.95]{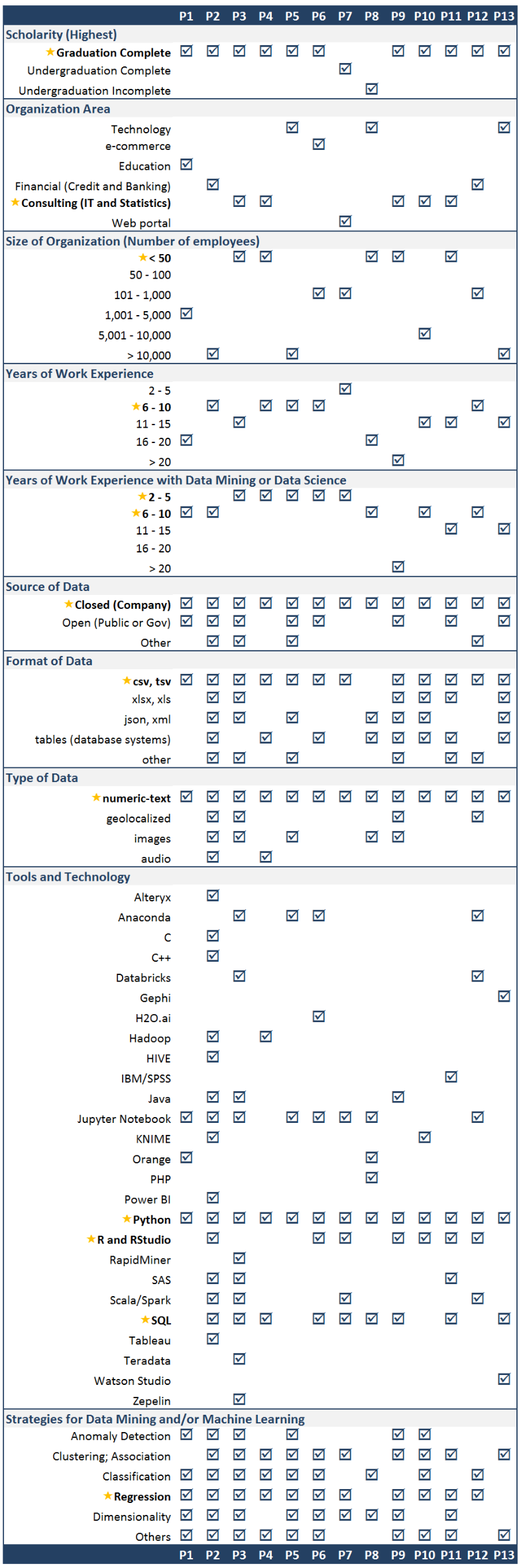}
    \caption{Additional information on the profile of the thirteen participants.}
    \label{fig:interview_profiles}
\end{figure}

The length of experience of the participants in the technology field ranged from 6 to 15 years and, with regards to data exploration more specifically, the range was reduced to 2 to 10 years. That happened because 62\% of the participants started working in positions outside data mining. Further details on the participants' profile is shown in Figure~\ref{fig:interview_profiles}.

\subsection{Procedure}

Each participant was interviewed continually, and the sessions lasted from 30 to 60 minutes. The same environment configuration was used for all participants, face-to-face or online conversations, i.e., calls or video conferences. First, we introduced the procedure and presented the consent form, in compliance with our Research Ethics Committee (REC). Subsequently, we briefly introduced our study and we provided participants with the opportunity to ask any questions regarding the explained items.

The interview was guided by a semi-structured questionnaire consisting of five parts and a total of 25 questions.
A copy of the questionnaire was shared with the participants during the interview. Additionally, we asked participants to consider their most recent data analysis projects while answering the questions.

A pilot interview was run to confirm the clarity of the questions and the approximate duration required for the activity. Since it occurred as planned, the content of the pilot interview was regarded as part of this study, as participant number 1. The interviews were performed in May, June, and July 2018, by the same interviewer. During each session, the interviewer took extensive notes of the answers. Parts of the sessions were recorded, with the consent of participants, and the audio was used to review the notes.

We developed the analysis code of the responses primarily following the same structure used for the questionnaire, divided into five parts. Afterwards, the questions related to each part worked as a second level of coding. We tabulated the collected data following these two levels, which resulted in 325 entries, i.e., each entry is the transcript for the open responses provided by each of the thirteen participants. In more details: Part 1, Participant Profile, resulted in 117 entries since there were nine questions; Part 2, Data Profile, resulted in 52 entries since there were four questions; Part 3, Data Analysis Process, resulted in 52 entries since there were four questions;
Part 4, Preprocessing Activities, resulted in 52 entries since there were four questions; Part 5, Visualization Techniques, resulted in 52 entries since there were four questions. Later, the content of each question was analyzed, comparing the responses of all participants. During that step, the third level of code was created to group similar responses. In the next subsection, we describe the recurring patterns and the significant elements observed during this analysis. As a rule, we considered the items reported by more than two participants. However, those items emphasized as important, even if only by one participant, were discussed as well.

\subsection{Analysis of the Interviews and Results}

The results and discussion based on the analysis of the responses were grouped into four items: data profile, data analysis process, preprocessing activities, and visualization of data quality issues.
The most relevant aspects are described in the following paragraphs. In relation to the numerical computation in this analysis, it is important to note we are only counting explicit responses. Therefore, for some situations, we cannot assume the other participants agree or disagree with a particular point since their answers were not counted.

\subsubsection{Data profile}

The information captured about the source, format, and type of data is summarized as part of Figure~\ref{fig:interview_profiles}. Regarding the volume of the datasets in use, it ranges from a small number of data records, i.e., which can be processed in simple spreadsheet, to Big Data~\cite{AIP_DEMAURO_2015} infrastructures, with billions of records and more than 100 thousand features.

\subsubsection{Data Analysis Process}

\begin{figure*}[!htb]
   \centering
   \includegraphics[scale=1]{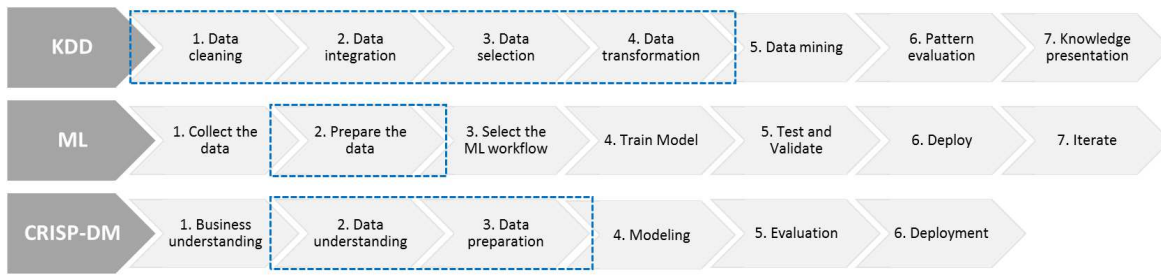}
    \caption{Example of workflows used during data analysis. The steps highlighted by the dashed box are typically be considered as part of the preprocessing phase.}
    \label{fig:workflows}
\end{figure*}

Participants described their work process similarly to KDD, Machine Learning (ML), or CRISP-DM workflows, see Figure~\ref{fig:workflows} for details. Moreover, the participants mentioned that the steps may vary according to the scope and type of project. For some cases, these workflow tasks were mixed, for instance, \textit{1. Business understanding} and \textit{2. Data understanding} from CRISP-DM were added as pre-steps in the KDD and ML workflows. One participant added a new step \textit{0. Research}, in order to represent the literature review in the domain under analysis, including model evaluations, prior to starting any other regular step.

When asked about the activities that usually require the most investment of time or that cause the most difficulties during execution, the reference to the preprocessing phase was almost unanimous. As reasons for that, they mentioned: bad quality of the data, lack of data standardization, infrastructure limitation, and mainly the efforts to understand the raw data prior to deciding on any transformations, for instance, data cleaning or the creation of new features. However, for three participants, the preprocessing stage was not highly demanding.

One works with Deep Learning with images, and their cycle started directly on \textit{3. Select ML algorithm} and \textit{4. Train model}, in reference to the ML workflow. The second considered \textit{1. Business understanding} and \textit{2. Data understanding}, in reference to CRISP-DM, more demanding. That occurred because they were developing a new solution and were not following the same structure of on-demand projects as most of the other participants. The third worked in a new organization that provides financial services; the company invested in its system architecture since the conception, leading to few data issues and no need to integrate with legacy systems.
 
Business understanding was the second task indicated as highly demanding because it requires domain expertise and, in some cases, the clients do not know what to ask or look for in their own data. Other items were also mentioned, such as data collection in the case of heterogeneous and complex systems and model deployment in the production environment.

Regarding their data mining strategies, the most indicated were Clustering, Association, Classification, and Regression Analysis. Additionally, many participants mentioned the dimensionality reduction strategy used as part of preprocessing. One participant said that for their context this was not a good strategy, and explained that if there are 300 attributes reduced to 10 dimensions, it will be necessary to guarantee all the 300 attributes arrive with quality in the production environment. Then, keeping the model working as planned after deployment adds more complexity to the process. Thus, they preferred to invest in a strategy that only selects the really important attributes. Furthermore, Principal Component Analysis (PCA) was indicated as still useful, but only with the purpose of understanding which attributes are interesting and should be kept, and not with the intention of working with dimensionality reduction in later stages.

\subsubsection{Preprocessing activities}

Nine participants reported preprocessing activities as laborious since they require a lot of manual intervention. Therefore, they were indicated as highly dependent on professional experience and domain expertise. Although they had already created a particular toolbox of strategies and scripts to make this process easier, the majority of the situations still requires the development of customized scripts to be aligned to the reality of their projects. In this context, Python~\cite{online_python} and R~\cite{online_r} play an important role. Four participants mentioned using tools such as Databricks~\cite{online_databricks}, KNIME~\cite{ACM_Berthold_2009, online_knime}, Gephi~\cite{bastian_2009, online_gephi}, and Orange~\cite{JMLR:demsar13a, online_orange} in some moments to support this process. Only one participant said that most of the preprocessing activities were performed directly on Spark~\cite{XXX_Zaharia_2010, online_spark}.

When asked to share further details about the preprocessing tasks, most described, or even emphasized, the following three activities. It is important to notice that the order of each activity is not the same for all participants and may vary according to their project engagement.

\textbf{1. Analysis.} Some participants considered a period of time to conduct an assessment of the business area to understand the problem and the data, especially when a domain expert was not involved. They described performing an exploratory analysis of raw data using statistical methods to generate data summaries. Subsequently, behaviors and distributions of these data were evaluated and the next activities were decided based on that. The understanding of how the variables are related was also considered within this exploratory analysis. Another item mentioned was the strategic plan to clean and standardize the data.

\textbf{2. Cleaning and standardization of data.} Most participants described performing the general cleaning of the data, trying to ensure the variables are from the same type, and other standardizations, e.g., data transformation to match the syntax rules defined by the database where newly arrived data is being appended. Additionally, few participants reported investing more time in the treatment of missing values, since there is the need to understand, for example, if they are system errors or forms where people do not need to fill in that information or even if they result from an incorrect cross-over during data collection. One participant classified this activity as data enrichment, which could be considered a part of the data quality process.

\textbf{3. Feature selection.} They reported evaluating the variables that may be interesting for the model and, from those, deciding the new variables to be created. In addition, some participants indicated they spent considerable time in this activity of categorical variable definition. One participant cited as an example that the cardinality of the variables could be a problem. Since sometimes it has a huge number of domains, and if, as a strategy, this variable is opened in flags, then soon there would be a lot of new flags that require treatment, leading to extra complexity. Thus, they indicated the need to be careful to understand which technique is going to be selected for each type of variable being treated. 

Additional challenges and frequent problems were indicated while describing their preprocessing efforts. The next items summarize them.

\textbf{Data volume and high dimensionality.} Opposite realities were reported: first, a group with a large volume of data and several attributes, e.g., 500 thousand columns in a table, where such high dimensionality becomes a challenge. On the other side, there were participants who noticed insufficient data, e.g., not a minimum number of records to conduct the analysis safely.

\textbf{Processing time.} Three participants reported some issues with their technical resources, which eventually became the bottleneck for some projects due to waiting time to process their data.

\textbf{Access to the data.} Another point mentioned was the difficulty to access the data, due to data confidentiality restrictions, owing to particularities of the businesses, such as financial services and healthcare.

\textbf{Data quality.} Eight participants considered data quality a frequent point of concern. Regarding the most frequent issues, the number one, mentioned by 92\% of participants, was Missing Values (Null/Empty), followed by Missing Records (69\%), Inconsistency-Ambiguous data (62\%), and Incorrect Issues, such as Duplicates (54\%) and Outliers/Non-Standard (54\%). Additionally, two participants indicated that the raw data always has problems, such as missing data and outliers. Hence, their starting point is looking for these issues. When they are not present, they then continue the investigation drilling down the specific variable to better understand its behavior. They emphasized this process as very dependent on the knowledge of the analyst performing the activity. On the other hand, three participants recognized they ignore some type of errors, such as Incorrect-Duplicated and Inconsistent-Ambiguous data, depending on the scope of the project and the volume of data.

\subsubsection{Visualization of Data Quality}

The beginning of the final part of the questionnaire related to the previous question on data quality issues but focused on how the participants notice these issues. The idea was to acquire further information on the visual identification of data issues, which could be used as a guideline during the development of new visualization techniques. However, when working with the text-numeric type of data, all participants reported the use of scripts to perform the data analysis, e.g., generation of the total count of Null per column. Hence, most of them relied primarily on the validation of the absolute numbers, based on their script outputs, rather than on visual exploration or use of any visualization techniques in the process. For unstructured data, e.g., audio and images, the participants mentioned the need for a manual inspection. 

When using visualization to support their analysis, they mentioned generating graphics such as bar plots, lines, radar plots, box plots, scatter plots, and histograms, which are available in visualization libraries for Python, e.g., matplotlib~\cite{online_matplotlib} and seaborn~\cite{online_seaborn}, and R, e.g., ggplot2~\cite{online_ggplot2}. In order to identify outliers, four participants indicated that boxplot could help to visualize the distribution. Other five participants mentioned the use of additional resources, such as the visualizations available on Hadoop~\cite{online_hadoop}, Orange, Gephi, Databricks, and KNIME.

Five participants emphasized that missing data was the most common problem related to data quality. In addition, they mentioned that tools like SAS~\cite{online_sas} can help with the identification of the missing data and even perform transformations automatically. Nevertheless, the solution to this problem cannot be seen so simply, and the validation of these transformations still requires manual inspection. In these cases, one participant said that first they used VIM~\cite{JSS_KOWARIK_2016, online_vim}, a graphical user interface available as an R package, to build visualizations to help understand the patterns of these missing values or \textit{NA}s, which stands for Not Applicable, Not Available, or Not Announced. 

So we could ask ourselves, what is the reason for them not to use, or use very little, visualization techniques during the process? Three participants argued that it occurs because they were dealing with a very large volume of data, which results in difficulties to visualize the data. Additionally, after the solution deployment, the preprocessing must be automatized and cannot be dependent on any manual intervention in the production environment. Then, a visualization could be used only during the initial problem analysis and for model changes. Other three participants mentioned that the choice related to the capacity of the current tools to handle data processing. Free tools, e.g., Orange, cannot process huge volumes, being valid only for proof of concept purposes. One participant observed that even tools that promise to handle Big Data, e.g., Gephi, did not do that in their experience. Moreover, one participant highlighted that even for the most robust tools, which could handle graphic rendering, it was still hard to capture any meaningful information from a crowded visualization if there was too much data.

Additionally, five participants stated that generating the visualization was time-consuming. Thus, due to the timeline of the projects, they preferred to invest their time in other activities and then only generate the final visualization that would be shared with the business team and/or clients. One participant also said their current scripting approach, which allowed to look directly at the numbers, was enough, which means there was no need to add any visualization technique during their analysis. Another participant mentioned that they did not know how to use visualization to support preprocessing activities, demonstrating a lack of communication between the visualization research community and the professionals of the enterprise.

In conclusion, the participants were encouraged to mention any visualization techniques or additional features to their current tools that could support their preprocessing activities. Their \textit{wishlist} was considered to build the ten insights introduced in the next section.

\section{Insights for New Visualizations}
\label{challenges_trends}

During our interviews, only one participant mentioned visualization was not a differential for the activities they were performing during preprocessing. Two other participants expressed they felt confident with their set of tools. However, the ten remaining participants demonstrated an interest in different ways to explore their data with visualization techniques. Based on these feedbacks and complementary to the discussion started in the previous sections, in this section, we present a list of ten insights for visualization in data exploration.

\begin{figure*}[!htb]
   \centering
   \includegraphics[scale=1]{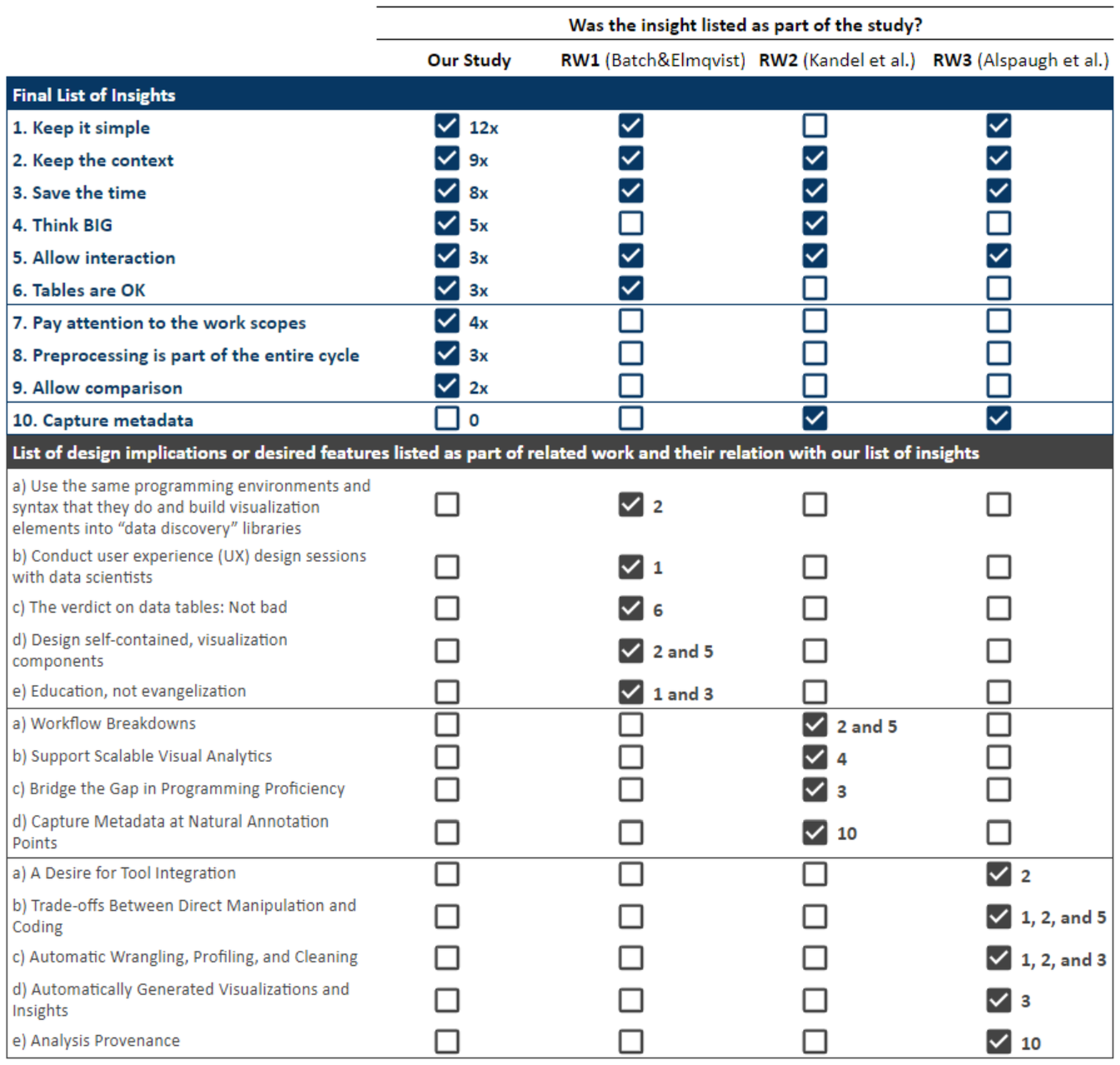}
    \caption{Complete list of the insights. (Top of figure, dark blue box) We present the final list of insights, their frequency in our study, i.e., how many participants mentioned it, and their connection with other studies. (Bottom of figure, gray box) We present the list of design implications or desired features we could identify in the three related works, and their relation to our final list of insights, indicated by the number of the insight.}
    \label{fig:complete_list_insights}
\end{figure*}

We compiled the final list of insights following an iterative, incremental coding method, which we explain in the next six steps: (1) first, the list started based on the inputs received from participant one while explaining his \textit{wishlist}. (2) Every input from a new participant was considered to review the latest version of the list, checking for similarities and complementing the background of the existing items or adding new items to the list. (3) After the completion of the interviews, all the records of the responses were reviewed, including all prior entries, to evaluate if any other item could be added based on the most common inputs, primarily related to challenges and improvement opportunities while describing any particular activity. (4) The items were labelled and ordered from the most to the least frequent. The items that were not mentioned by at least two participants were not included in the final list. (5) We merged the list of recommendations for tool development or design implications available in the related works with the list obtained in Step 4, which resulted in one additional insight. (6) Finally, we ordered the list considering Step 4 for the insights in common with the related work, i.e., from Insight 1 to 6, then the insights that were only identified in our study, i.e., from Insight 7 to 9, and lastly the additional insight not covered by our interviews, i.e., Insight 10. In Figure~\ref{fig:complete_list_insights}, we added details on the list of insights and the correlation of each source that mentioned them.

To simplify the description of the comparison with the related works, we will continue using the following code: RW1 for Batch and Elmqvist~\cite{IEEE_BATCH_2018}, RW2 for Kandel et al.~\cite{IEEE_KANDEL_2012}, and RW3 for Alspaugh et al.~\cite{IEEE_Alspaugh_2018}.

\subsubsection{Keep it simple}
For the majority of the cases, the existing visualizations or more traditional charts should fulfill the demand, without the need for novel visualization techniques, but rather focusing on reusable artifacts and recommendation features according to the type of data and what is intended to be presented. Moreover, even though Python's and R's current visualization packages and libraries are easy to use, they still require some level of programming. Hence, a more ready-to-play alternative, such as Tableau~\cite{online_tableau} and Qlik~\cite{online_qlik}, but easier to use, could encourage the use during the preprocessing phase instead of just at the end of the process. The perception that traditional charts are considered good was only stated by RW1. Moreover, RW1 noticed a lack of usability attention for visualization solutions applied to data mining. Therefore, user experience (UX) design sessions were indicated, and this can support to keep the solution simple for real scenarios use. However, only RW3 objectively mentioned the need for easier tools as desired by data analysts.

\subsubsection{Keep the context}
Any new solution should remain compatible with the most used tools for data mining, currently Python and R, in order to build an uninterrupted work environment, preventing data analysts from losing the context under investigation while alternating among several different tools. Complementary, RW1 stated it is important to keep the same syntax of the programming environments used by data analysts. In addition, it indicated the relevance of considering the integration with command line interfaces and of building ``visualization elements into data discovery libraries''. Although RW2 did not objectively mention it as part of the programming environment, this paper referred to the need for visualization tools to avoid the breakdown of the workflows, hence, directly promoting connections to the existing environments. The same was indicated by RW3, which is not focused on the visualization features, but was considered important for data exploration tools as a whole. 

In addition, new tools should allow the evaluation of multiple rows and attributes on the same view, without losing the context under investigation. Thus, there is a need to plan the use of interaction techniques such as \textit{focus+context}, where ``a selected subset of the structure (focus) is presented in detail, while the rest of the structure is shown in low detail to help the viewer maintain context'' \cite{Book_Ward_2015}, therefore avoiding the \textit{change blindness} effect, related to the difficulty to notice changes made during an eye movement~\cite{ELSEVIER_RENSINK_2000}.  

\subsubsection{Save the time}
Complementing the previous point, the new visualization tools should consider intuitive features and little need for configuration and/or coding, aiming to keep the agility in the working process. Data analysts also regarded the visualization as ``too time-consuming to be worth their efforts'' during the discussion in RW1. The same was observed in RW3, where the data analysts expressed difficulties around visualizations, such as choosing the right type of chart. Similarly, RW2 discussed this idea as required to ``bridge the gap in programming proficiency'', since most of the professionals without ``hacker'' skills, per their study classification, faced difficulties to manipulate data from diverse sources and especially during the wrangling tasks. Thus, a solution that automatically generates some examples or basic templates to support its use and provides recommendations of visualization techniques based on the type of data could be very useful. As a consequence, this approach should avoid some unsuitable uses, such as the use of bar charts expecting to see trends, when they are better to evidence volumes.

\subsubsection{Think BIG}
New visualizations should support scalable solutions, considering Big Data needs. Data rendering can be very difficult, even to plot simple scatter plots when dealing with large volumes of data. However, it is a growing demand and the development of techniques that can handle this scenario is urged, such as the ones using density or aggregation plotting, even though it requires different strategies, such as data reduction by selecting a sample and server-side preprocessing, to be explored. The same was discussed in RW2 under the statement ``scaling visualization requires addressing both perceptual and computational limitations''. RW2 was published in 2012, and this subject remains a critical challenge.

\subsubsection{Allow interaction}
It is important to provide more than static reports. Moreover, allowing the data analyst to perform flexible data manipulation within visualization tools is fundamental. RW1 indicated the visualization components should enable full-fledged interaction, such as zooming and panning, filtering, and details on demand~\cite{IEEE_Shneiderman_1996}. It is aligned with the techniques suggested by us in insight 1, \textit{Keep the context}. As an example, one participant mentioned that a solution similar to Orange UI's proposal, but in a more robust and online version, could contribute to filling this gap, while for RW3 ``embedding interactive visualizations within notebook-style'' is a better approach considering the emerging trends.

\subsubsection{Tables are OK}
The tabular format is considered a good choice for visual representation. The same was noticed in RW1. Files to store tabular data and structured database tables are widely used. However, there are still opportunities to be explored for table views, such as combining different interaction options and visualization techniques like Table Lens~\cite{ACM_Rao_1994} or Pixel-oriented~\cite{IEEE_KEIM_2000}.

\subsubsection{Pay attention to the work scopes}
Two work scopes were indicated during our interviews as lacking attention by current visualizations solutions, which remains a good opportunity for future works. One concerns the creation of new variables, features, which usually requires a lot of analysis time during preprocessing activities. The other is related to the deep learning scope for visual interpretation of why each decision was made. In addition, more interactive visualizations to support the parameterization options are needed.

\subsubsection{Preprocessing is part of the entire cycle}
For many data mining workflow processes, such as Visual Analytics~\cite{book_Keim_2010} and KDD~\cite{Morgan_2011_Han}, preprocessing is represented as part of a flow in a one-way direction, similarly to a waterfall approach. However, we could notice during the interviews that for most cases multiple interactions were required among preprocessing activities and all the other stages during the same cycle. Except for confirmatory analysis, where most of the process was already automated and little interaction was needed, for other cases, especially for initial data exploration, multiple back and forwards in the raw data occurred. This matches with the progressive paradigm that enables the data analyst to inspect partial results as they become available and interact with the algorithm to prioritize items of interest, as explained by Stopler et al.~\cite{IEEE_Stolper_2014} while introducing the Progressive Visual Analytics.

\subsubsection{Allow comparison}
Considering adding features that allow the comparison of data prior to and after its transformation is important to support the preprocessing decision. 
It could follow a similar approach as proposed by Kindlmann and Scheidegger~\cite{IEEE_Kindlmann_2014}, which discussed the importance of knowing whether data transformations respected the original data. Furthermore, one participant mentioned that despite preprocessing activities being very fundamental and at some level performed by all data analysts, few people are truly proficient at them. Hence, this visual support could contribute for more data analysts to adopt visualization as part of their daily strategies, since most of them complained about the difficulties during data cleaning or wrangling activities.

\subsubsection{Capture Metadata}
Besides the two previous insights, if automatic exploratory tasks or data transformations are needed, it is important to present the logic underneath them, because, as identified by RW2 and RW3, data analysts desired to continue working with control and visibility of what the tool was doing. Thus, the creation of metadata for the dataset under analysis is fundamental to this process.

\section{Discussion and Limitations}
\label{discussion_limitations}


With respect to opportunities for improving our study, we can list two main items: first regarding to the procedure. The number of questions was designed to guarantee that each interview session would take no longer than one hour, in an attempt to capture a higher number of positive returns to our participation invitation. However, a more open strategy for data collection such as an experiment where participants are instructed to perform a list of tasks and it is possible to observe how they deal with them to solve certain problems, could contribute to acquire further details about daily practices. Likewise, that approach would require an additional number of hours, at least two hours for each participant session based on RW1 study, and possibly reducing the list of participants available to join the activity.

The second opportunity is regarding the participant's profile. Most of our interviewees were working in the IT Industry. Additional participants from different organization structures, such as government, could contribute to a different perspective. Also, we notice lack of female representation, but that seems to be a bigger issue in the STEM (science, technology, engineering, and mathematics) areas. 
Therefore, despite our efforts to recruit a variety of participants, the data collected and its analysis cannot be considered a representation of all data analysts.

The last insight presented in our list, \textit{10. Capture Metadata}, was the only one seen in the related works that was not captured during our interviews. On the other hand, the insights \textit{7. Pay attention to the work scopes}, \textit{8. Preprocessing is part of the entire cycle}, and \textit{9. Allow comparison} in our list were not mentioned by any of the indicated related works, which brings new topics for discussion. Moreover, none of the other insights appeared together in the final list of recommendations or implications for design, as shown in Figure~\ref{fig:complete_list_insights}. 

Although RW1 was very well organized, introducing relevant points to this discussion, an important item related to the need for scalable solutions, insight \textit{4. Think BIG}, was not listed in its final implications for design. Similarly, despite RW2 being one of the first studies addressing this subject and reporting important perceptions from enterprise data analysis, it still did not cover our entire list, nor did it present its design implications in an approach that is as straightforward as ours. Besides, it was not concerned with the particular needs of data mining. While RW3 also contributed with this discussion, their primary focus was neither visualization nor preprocessing activities in data mining. Thus, many of its recommendations covered data exploration at a higher level of the process than ours.

\begin{figure}[!htb]
   \centering
   \includegraphics[scale=1.1]{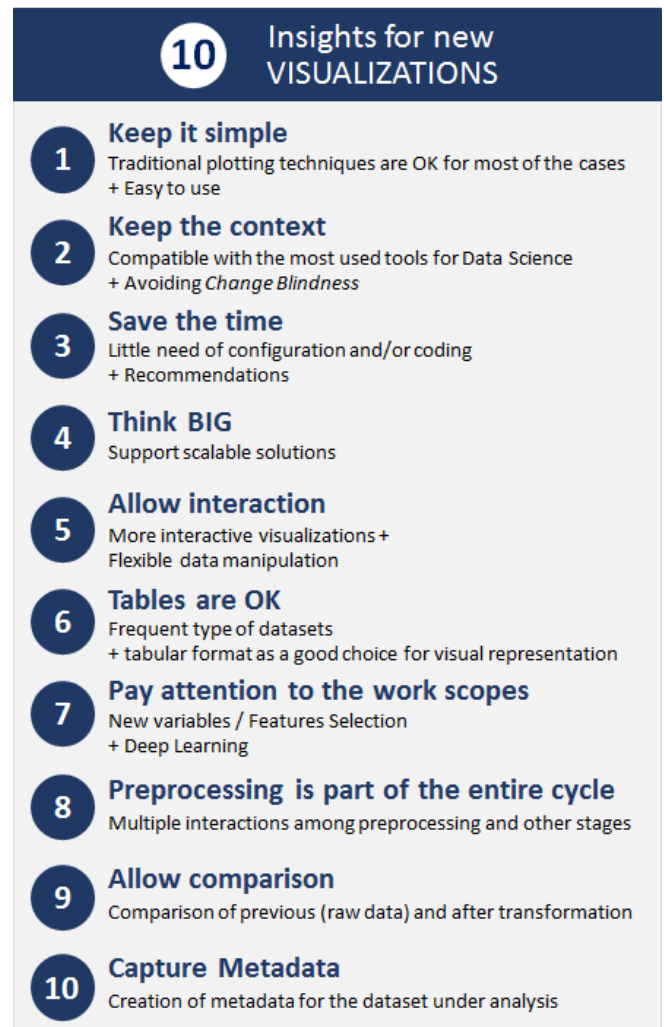}
    \caption{Consolidated list of insigths for new visualizations solutions.}
    \label{fig:insights}
\end{figure}

As summarized in Figure \ref{fig:insights}, we hope to contribute with a straight and easy-to-understand list of items that require attention when planning new visualization solutions as part of the alternatives to lower adoption barriers. Moreover, despite our focus on the preprocessing phase for many of our questions, we consider these insights are also applicable to other phases of the data mining workflow, which includes the final visualizations used to report the analysis and findings.

\section{Conclusion and Future Work}
\label{conclusion}

We interviewed thirteen enterprise professionals to understand their data analysis practices in data mining, how they use visualization during the preprocessing phase, and which features could support them during this process. Additionally, we presented the methodology used for data collection in this interview study and the results obtained from the interviews.

Our main contribution was the organization of the challenges and opportunities identified during our analysis of the interviews, which resulted in a list of ten insights. This list of insights was then compared with the closest related works, improving the reliability of our findings, and, at the same time, encouraging the discussion about uncovered considerations.

Even though some insights appeared in previous studies, an in-depth analysis of the related works was necessary to identify and relate their findings to our final list of insights. Through our study, we also summarized practical items to be considered during the planning and development stages of new visualization solutions, aiming to lower the barriers to adopt visualization as part of any data mining workflow. Ultimately, this study contributes as a source of requirements to fill the visualization gap during the initial exploratory analysis.

While contemplating the requirements elicited by our study, several future work opportunities arise. To begin, we plan to develop preliminary prototypes considering our list of insights. To conclude, we intend to evaluate the prototypes while conducting in-depth interviews or user-centered experiments with the participation of domain experts.

\section*{Acknowledgment}
This study was financed in part by the Coordena\c{c}\~{a}o de Aperfei\c{c}oamento de Pessoal de N\'{i}vel Superior – Brasil (CAPES) - Finance Code 001. Also, in cooperation with HP Brasil Ind\'{u}stria e Com\'{e}rcio de Equipamentos Eletr\^{o}nicos LTDA. using incentives of Brazilian Informatics Law (Law number 8.2.48 of 1991). Finally, with the support of the Government of Canada.

\bibliographystyle{SageV}
\bibliography{main.bib}

\end{document}